\documentclass[aps,prc,twocolumn,nofootinbib,superscriptaddress,amsmath,amssymb]{revtex4-1}

\usepackage{graphicx}
\usepackage{bm}
\usepackage{hyperref}
\usepackage{dcolumn,dsfont}
\usepackage{color}
\usepackage{xspace}

\renewcommand{\vec}[1]{\mathbf{#1}\xspace}

\newcommand{\MeV}{\,\mathrm{MeV}\xspace}

\newcommand{\fm}{\,\mathrm{fm}\xspace}
\newcommand{\fmi}{\,\mathrm{fm}^{-1}\xspace}
\newcommand{\fmiq}{\,\mathrm{fm}^{-3}\xspace}
\newcommand{\emax}{$e_\text{Max}$\xspace}
\newcommand{\hw}{$\hbar \omega$\xspace}

\begin{document}

\title{Probing chiral interactions up to next-to-next-to-next-to-leading order in medium-mass nuclei}

\author{J.\ Hoppe}
\email[Email:~]{jhoppe@theorie.ikp.physik.tu-darmstadt.de}
\affiliation{Institut f\"ur Kernphysik, Technische Universit\"at Darmstadt, 64289 Darmstadt, Germany}
\affiliation{ExtreMe Matter Institute EMMI, GSI Helmholtzzentrum f\"ur Schwerionenforschung GmbH, 64291 Darmstadt, Germany}

\author{C.\ Drischler}
\email[Email:~]{cdrischler@berkeley.edu}
\affiliation{Department of Physics, University of California, Berkeley, CA 94720}
\affiliation{Lawrence Berkeley National Laboratory, Berkeley, CA 94720}

\author{K.\ Hebeler}
\email[Email:~]{kai.hebeler@physik.tu-darmstadt.de}
\affiliation{Institut f\"ur Kernphysik, Technische Universit\"at Darmstadt, 64289 Darmstadt, Germany}
\affiliation{ExtreMe Matter Institute EMMI, GSI Helmholtzzentrum f\"ur Schwerionenforschung GmbH, 64291 Darmstadt, Germany}
 
\author{A.\ Schwenk}
\email[Email:~]{schwenk@physik.tu-darmstadt.de}
\affiliation{Institut f\"ur Kernphysik, Technische Universit\"at Darmstadt, 64289 Darmstadt, Germany}
\affiliation{ExtreMe Matter Institute EMMI, GSI Helmholtzzentrum f\"ur Schwerionenforschung GmbH, 64291 Darmstadt, Germany}
\affiliation{Max-Planck-Institut f\"ur Kernphysik, Saupfercheckweg 1, 69117 Heidelberg, Germany}

\author{J.\ Simonis}
\email[Email:~]{simonis@uni-mainz.de}
\affiliation{\mbox{Institut f\"ur Kernphysik and PRISMA Cluster of Excellence, Johannes Gutenberg-Universit\"at, 55099 Mainz, Germany}}

\begin{abstract}
We study ground-state energies and charge radii of closed-shell medium-mass 
nuclei based on novel chiral nucleon-nucleon (\textit{NN}) and three-nucleon (3\textit{N}) interactions, 
with a focus on exploring the connections between finite nuclei and nuclear matter. 
To this end, we perform in-medium similarity renormalization group (IM-SRG) 
calculations based on chiral interactions at next-to-leading order (NLO), N$^2$LO, 
and N$^3$LO, where the 3\textit{N} interactions at N$^2$LO and N$^3$LO are fit to 
the empirical saturation point of nuclear matter and to the triton binding energy. 
Our results for energies and radii at N$^2$LO and N$^3$LO overlap within 
uncertainties, and the cutoff variation of the interactions is within the EFT uncertainty 
band. We find underbound ground-state energies, as expected from the comparison 
to the empirical saturation point. The radii are systematically too large, but the 
agreement with experiment is better. We further explore variations of the 3\textit{N} couplings 
to test their sensitivity in nuclei. While nuclear matter at saturation density is quite 
sensitive to the 3\textit{N} couplings, we find a considerably weaker dependence in 
medium-mass nuclei. In addition, we explore a consistent momentum-space 
SRG evolution of these \textit{NN} and 3\textit{N} interactions, exhibiting improved many-body 
convergence. For the SRG-evolved interactions, the sensitivity to the 3\textit{N} couplings 
is found to be stronger in medium-mass nuclei.
\end{abstract}

\maketitle

\section{Introduction}
\label{sec:intro}

The development of improved nucleon-nucleon (\textit{NN}) and three-nucleon (3\textit{N}) interactions 
within chiral effective field theory (EFT) for \textit{ab initio} studies of atomic nuclei and 
infinite nuclear matter is currently a very active field of
research~\cite{Ekst15sat,Carl15sim,Lynn16QMC3N,Rein17semilocal,Ente17EMn4lo,Dris17MCshort}.
While none of the presently available interactions is able to simultaneously describe
experimental ground-state energies and charge radii of nuclei over a wide
range of the nuclear chart, recent calculations not unexpectedly indicate a 
strong correlation between predictions for medium-mass nuclei and nuclear matter
properties~\cite{Hebe11fits,Ekst15sat,Simo16unc,Hage16NatPhys,Simo17SatFinNuc,Ekst17deltasat}.
Although a systematic and quantitative understanding of this correlation is still
missing, these studies highlight the significance of realistic saturation
properties for the construction of next-generation nuclear forces. In
particular, calculations based on interactions fitted only to \textit{NN} and few-body
observables tend to exhibit significant deviations from experiment for
heavier nuclei~\cite{Bind14CCheavy,Carl15sim,Tich16HFMBPT}. Remarkably,
one exception to this general trend was found in Ref.~\cite{Hebe11fits}. In 
this work, a family of \textit{NN} plus 3\textit{N} interactions was constructed using similarity 
renormalization group (SRG) evolved \textit{NN} interactions combined with
the leading chiral 3\textit{N} forces fitted to the $^3$H binding energy and
the charge radius of $^4$He. While all constructed interactions lead to a
reasonable description of the empirical saturation point, the ground-state
energies for closed-shell nuclei ranging from $^4$He to $^{100}$Sn
are well reproduced for one particular interaction ("1.8/2.0"), whereas 
charge radii are somewhat too small~\cite{Simo17SatFinNuc,Morr17Tin}.
However, the physical reasons of this remarkable agreement with this 
specific \textit{NN}+3\textit{N} interaction is still an open question. 

These findings suggest incorporating experimental constraints of heavier
nuclei directly into the fitting process of nuclear interactions, as was done in
Ref.~\cite{Ekst15sat}. Naturally, results based on these
interactions show in general better agreement with experiment for medium-mass
nuclei, but \textit{NN} scattering phase-shifts can only be reproduced to rather low
energies, when considering chiral interactions up to next-to-next-to-leading 
order (N$^2$LO). In Ref.~\cite{Dris17MCshort}, a complementary strategy
was pursued to fit the 3\textit{N} low-energy couplings $c_D$ and $c_E$ for
fixed \textit{NN} interactions to the $^3$H binding energy and the saturation region
of nuclear matter using a novel many-body perturbation theory framework 
for nuclear matter. It was found that, based on the \textit{NN} interactions of 
Ref.~\cite{Ente17EMn4lo}, a reasonable reproduction of the saturation point
can be obtained for all interactions at N$^2$LO and N$^3$LO and for different 
cutoffs $\Lambda = 450$ and $500 \MeV$. In this work, we extend this
study to interactions with a lower cutoff $\Lambda = 420 \MeV$ and 
investigate in detail the properties of medium-mass nuclei based on these
interactions using the \textit{ab initio} in-medium similarity renormalization
group (IM-SRG). This work thus presents the first N$^3$LO calculations of
medium-mass nuclei.

The paper is organized as follows: We briefly discuss the employed chiral
\textit{NN} plus 3\textit{N} interactions in Sec.~\ref{sec:chiral_int} and the IM-SRG 
many-body calculations in Sec.~\ref{sec:imsrg}. In Sec.~\ref{sec:results}, 
we study the model-space convergence and present results for 
ground-state energies and charge radii of medium-mass nuclei up to
nickel isotopes. We study the effects of consistent momentum-space
SRG evolutions of the \textit{NN} plus 3\textit{N} interactions and investigate the 
systematics of the results for nuclei and matter with respect to variations
of the low-energy couplings. Finally, we conclude and give an outlook in
Sec.~\ref{sec:summary}.

\section{Chiral interactions} 
\label{sec:chiral_int}

\begin{table}[b]
\caption{\label{tab:cD_cE}
Three-nucleon couplings $c_D$ and $c_E$ for the N$^3$LO EMN \textit{NN} potentials 
420, 450, and 500~MeV~\cite{Ente17EMn4lo,Mach17privcom}, which reproduce the
experimental $^3$H ground-state energy $E(^3\text{H})=-8.482 \MeV$ at N$^3$LO.
The last row gives the results from simultaneous fits to the $^3$H binding energy and
the empirical saturation region of symmetric nuclear matter (see
Ref.~\cite{Dris17MCshort} for details for 450 and 500~MeV). The fits to the empirical 
saturation point for $\Lambda = 420 \MeV$ are shown in Fig.~\ref{fig:cD_cE_EMN400_420}.}
\begin{ruledtabular}
\begin{tabular}{cc|cc|cc}
\multicolumn{2}{c|}{420~MeV} & \multicolumn{2}{c|}{450~MeV} & \multicolumn{2}{c}{500~MeV} \\ 
\quad $c_D$ \quad & \quad $c_E$ \quad & \quad $c_D$ \quad & \quad $c_E$ \quad & \quad $c_D$ \quad & \quad $c_E$ \quad \\
\hline
\quad $-5.0$ \quad & \quad $-2.509$ \quad & \quad $-5.0$ \quad & \quad $-2.149$ \quad & \quad $-5.0$ \quad & \quad $-2.534$ \quad \\
\quad $0.0$ \quad & \quad $-1.558$ \quad & \quad $0.0$ \quad & \quad $-1.321$ \quad & \quad $0.0$ \quad & \quad $-1.848$ \quad \\
\quad $5.0$ \quad & \quad $-0.685$ \quad & \quad $5.0$ \quad & \quad $-0.636$ \quad & \quad $5.0$ \quad & \quad $-1.573$ \quad \\
\hline
\quad $4.0$ \quad & \quad $-0.853$ \quad & \quad $0.25$ \quad & \quad $-1.280$ \quad & \quad $-2.75$ \quad & \quad $-2.190$ \quad
\end{tabular}
\end{ruledtabular}
\end{table}

\begin{figure}[t]
\centering
\includegraphics[width=\columnwidth,clip=]{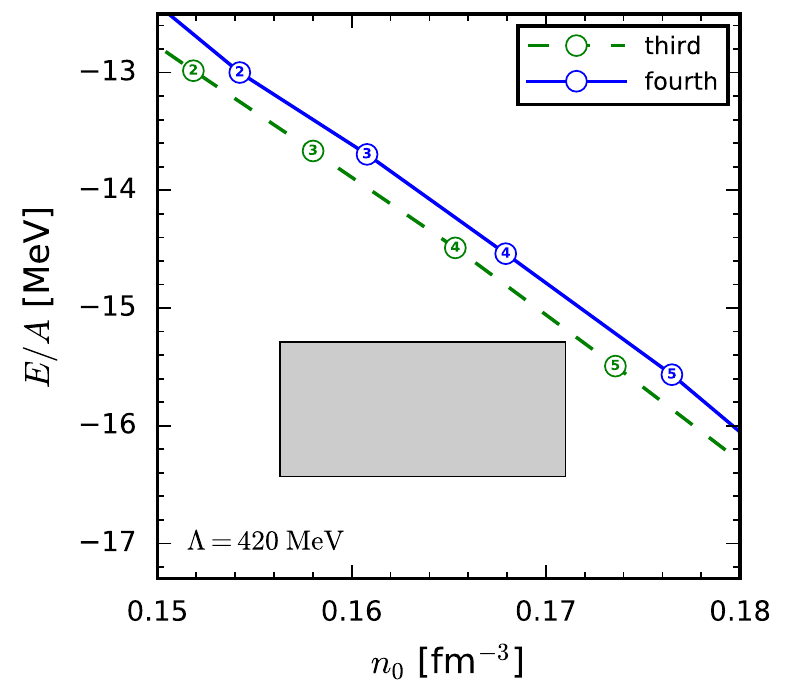}
\caption{\label{fig:cD_cE_EMN400_420}
Saturation point of symmetric nuclear matter at third (green-dashed line) and fourth
(blue-solid line) order in MBPT as trajectory of $c_D$ (see annotated values inside
the circles), while $c_E$ is determined by the $^3$H binding energy. The results are 
based on the N$^3$LO EMN $420 \MeV$ \textit{NN} potential with consistent 3\textit{N} forces at
N$^3$LO. For details on the fits to the empirical saturation region (gray box) see 
Ref.~\cite{Dris17MCshort}.}
\end{figure}

As in the recent N$^3$LO nuclear matter study~\cite{Dris17MCshort}, we use the
nonlocal chiral \textit{NN} interactions of Entem, Machleidt, and Nosyk (EMN)~\cite{Ente17EMn4lo}
with cutoffs $\Lambda = 450$ and $500 \MeV$. In addition, we consider a softer
N$^3$LO interaction with cutoff $\Lambda = 420 \MeV$~\cite{Mach17privcom}. The 3\textit{N} interactions
are regularized via a nonlocal regulator
\begin{equation}
\label{eq:3N_cutoff}
f_{\Lambda_{\text{3\textit{N}}}}(p,q) = \exp\left[-\left(\frac{4p^2 + 3q^2}{4 \Lambda_{\text{3\textit{N}}}^2}\right)^4 \right] \,,
\end{equation}
where $p$ and $q$ are the magnitudes of the relative momenta $\vec{p}$ and
$\vec{q}$, respectively~\cite{Epel02fewbody}, and (by choice) $\Lambda_{\text{3\textit{N}}} = \Lambda$.
In Ref.~\cite{Dris17MCshort}, we studied saturation properties of symmetric nuclear matter 
of these \textit{NN} potentials combined with consistent 3\textit{N} forces up to N$^3$LO using a new 
Monte Carlo framework that enables high-order calculations in many-body perturbation 
theory (MBPT). It was shown that fits to the triton binding energy and the empirical saturation 
point lead to narrow ranges for the two 3\textit{N} low-energy couplings $c_D$ and $c_E$ (note that
the fit to nuclear matter was not optimized to high accuracy). The resulting values of the 
3\textit{N} couplings are given in the last row of Table~\ref{tab:cD_cE}. In addition to 
Ref.~\cite{Dris17MCshort}, we explore the softer N$^3$LO interaction with $\Lambda = 420 \MeV$
for nuclear matter saturation in Fig.~\ref{fig:cD_cE_EMN400_420}. For this study, we take $c_D = 4.0$ 
(varied in steps of 1), to explore the reproduction of the saturation density.

In this work, we study the properties of medium-mass nuclei for the first time to N$^3$LO using
the IM-SRG framework. The relative 3\textit{N} matrix elements up to N$^3$LO have been 
calculated in Ref.~\cite{Hebe15N3LOpw}. We consider both unevolved and for the first time
consistently momentum-space SRG-evolved \textit{NN} and 3\textit{N} interactions following 
Ref.~\cite{Hebe12msSRG}. The SRG can significantly improve the rate of convergence 
of many-body calculations at the cost of induced many-body forces that may be sizable 
depending on the resolution scale of interest. Such induced contributions cannot be included
beyond the 3\textit{N} level at the moment, but the residual sensitivity of our results on the flow 
parameter serves as an estimate of the uncertainty due to neglected higher-body contributions.

\section{In-medium similarity renormalization group}
\label{sec:imsrg}

The IM-SRG takes advantage of normal ordering with respect to a chosen
reference state and decouples particle-hole excitations from the ground state
by a continuous sequence of unitary transformations to solve the
many-body problem~\cite{Tsuk11IMSRG,Herg13IMSRG,Simo17SatFinNuc,Herg16PR,Stro17ENO}. 
Similar to the free-space SRG~\cite{Bogn07SRG,Bogn10PPNP}, the flow 
equation for the Hamiltonian is given by
\begin{equation}
\label{eq:flow_eq}
\frac{\mathrm{d} H(s)}{\mathrm{d}s} = \left[\eta(s), \, H(s) \right] \,,
\end{equation}
with the flow parameter $s$ and the generator $\eta(s)$. For the IM-SRG, we take 
the arctan generator following the work of White~\cite{Whit02cantrans}.

The commutator relation~\eqref{eq:flow_eq} induces up to $A$-body contributions
in an $A$-body system. Including all of these induced terms is not feasible at the
moment, making a truncation scheme necessary. We use the IM-SRG(2), in which 
all operators are truncated at the normal-ordered two-body level and apply the 
Magnus formalism~\cite{Magn54exp,Morr15Magnus} to the flow equations instead
of an ordinary differential equation solver. By calculating the unitary transformation
underlying the IM-SRG directly, this approach is less memory demanding and faster,
especially for operators other than the Hamiltonian.

To calculate charge radii, we also evolve the intrinsic point-proton mean-square 
radius operator 
\begin{equation}
\label{eq:Rp2}
R_p^2 = \frac{1}{Z} \sum \limits_{i=1}^A \frac{1 + \tau_3^{(i)}}{2} \left(\vec{r}_i-\vec{R}\right)^2 \,,
\end{equation}
using the Magnus formalism, where $\vec{r}_i$ ($\vec{R}$) is the nucleon 
(nucleus center-of-mass) coordinate. $A$ and $Z$ are the mass and proton number,
respectively, and the operator $(1+\tau_3^{(i)})/2$ with the third component of the
isospin operator $\tau_3^{(i)}$, projects on protons. Taking into account the proton 
and neutron mean-square charge radius $\left<r_p^2\right>= 0.770$~fm$^{2}$ and 
$\left<r_n^2\right>=-0.1149$~fm$^{2}$~\cite{PDG16review},  as well as the 
relativistic Darwin-Foldy correction~\cite{Fria97foldyShift,Fold49nonrellim}
$3/(4m_p^2 c^4) = 0.033$~fm$^2$, and the spin-orbit correction
$\left<r^2\right>_{\text{so}}$~\cite{Ong10spinorbit}, we obtain the mean-square 
charge radius from
\begin{equation}
\label{eq:Rch_corrections}
R_{\text{ch}}^2 = R_p^2 +  \left<r_p^2\right> + \frac{N}{Z}\left<r_n^2\right> + \frac{3}{4m_p^2 c^4} + \left<r^2\right>_{\text{so}} \,,
\end{equation}
with neutron number $N$. For further details on the evaluation, we refer to 
the calculation of charge radii in the IM-SRG in Ref.~\cite{Simo17SatFinNuc}.

\section{Results}
\label{sec:results}

We first study the model-space convergence with respect to the harmonic-oscillator
single-particle basis with quantum numbers $e = 2n + l 
\leqslant$ \emax and oscillator frequency \hw. As usual, an
additional cut for the 3\textit{N} interaction matrix elements in the single-particle
basis is introduced by $e_1 + e_2 + e_3 \leqslant E_\text{3Max} < 3$\emax.
For the single-particle 3\textit{N} matrix elements, our results are based on Ref.~\cite{Simo17SatFinNuc}.
For the transformation of relative 3\textit{N} matrix elements to the single-particle basis,
we apply the truncation $\mathcal{J} \leqslant \mathcal{J}_\text{Max} = 9/2$
in the relative total three-body angular momentum $\mathcal{J}$ for unevolved
interactions with the relative total two-body angular momentum $J_{\text{max}} = 
8, 7, 6$ for $\mathcal{J} \leqslant 5/2$, $\mathcal{J} = 7/2$, and $\mathcal{J} = 9/2$,
respectively. For SRG-evolved interactions we use $\mathcal{J}_\text{max} =
7/2$ and $J_{\text{max}} = 5$.  Contributions to the ground-state energies
beyond these limits for these interactions are expected to be at the level of MeV,
which is small compared to the interaction sensitivity explored in the following.

\begin{figure}[t]
\centering
\includegraphics[width=\columnwidth,clip=]{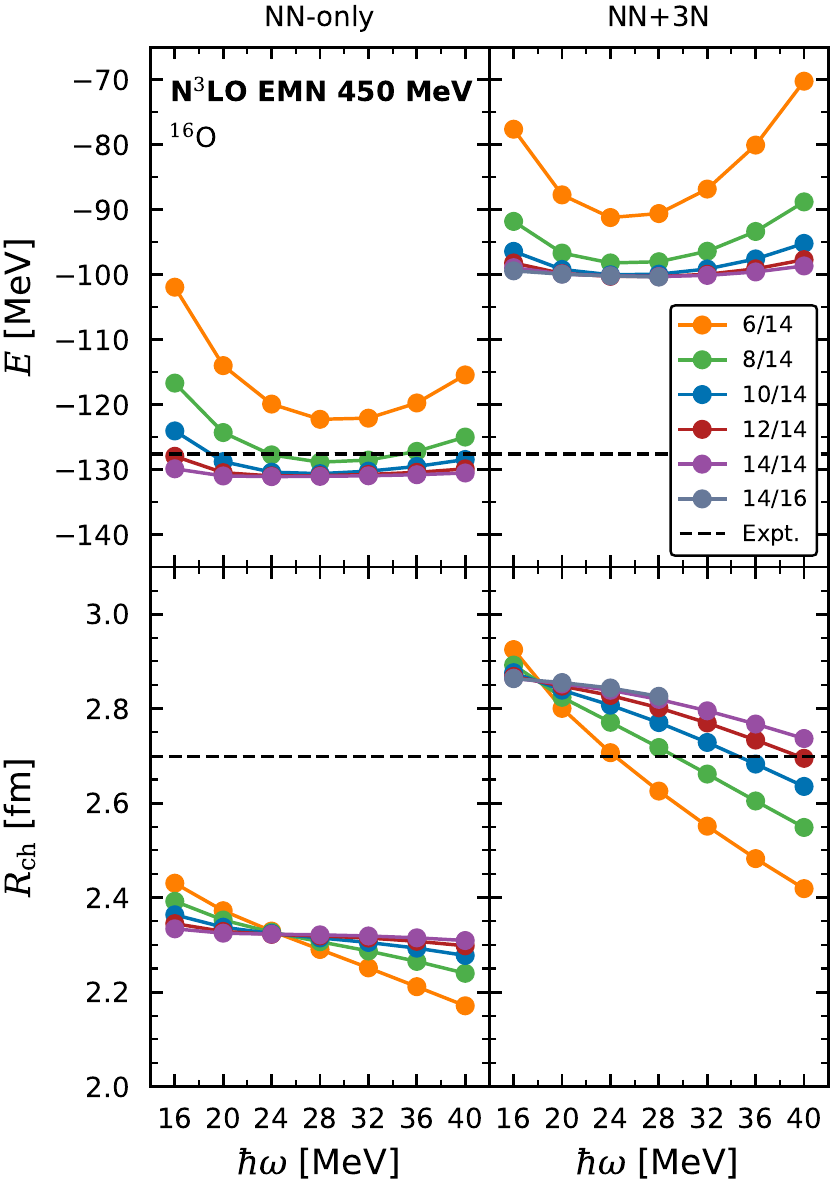}
\caption{\label{fig:Egs_Rch_O16_NN_3N}
Ground-state energy (top) and charge radius (bottom panel) of $^{16}$O as a function 
of the harmonic-oscillator frequency \hw at N$^3$LO for the \textit{NN}-only EMN $450 \MeV$
potential and the consistent \textit{NN}+3\textit{N} interaction in the left and right panels, respectively.
Results are shown for different sizes of the single-particle basis, \emax $= 6, 8, 10, 12,$ 
and 14, and for $E_{\text{3Max}}=14$ and 16, denoted by \emax/$E_{\text{3Max}}$. 
The experimental values (black-dashed lines) are taken from Refs.~\cite{Wang17AME16,Ange13rch}.}
\end{figure}

\begin{figure}[t]
\centering
\includegraphics[width=\columnwidth,clip=]{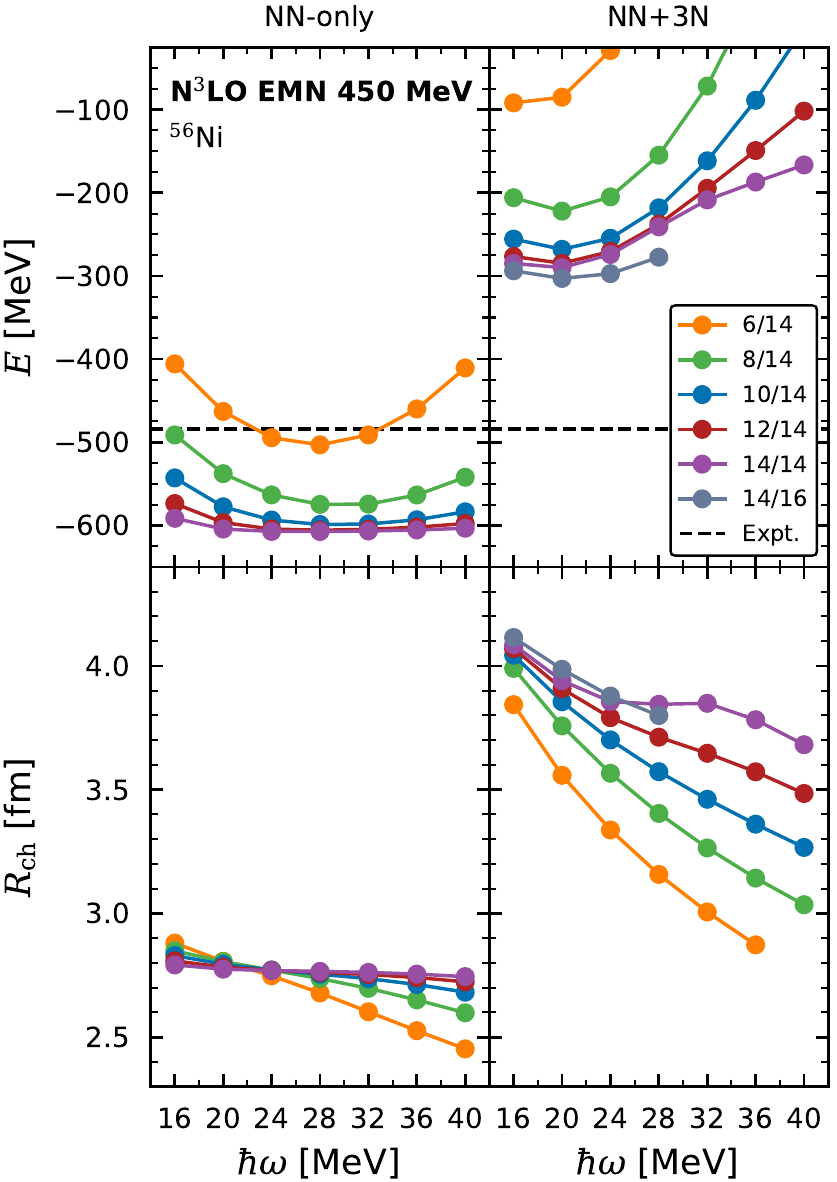}
\caption{\label{fig:Egs_Rch_Ni56_NN_3N}
Same as Fig.~\ref{fig:Egs_Rch_O16_NN_3N} but for $^{56}$Ni.}
\end{figure}

In Figs.~\ref{fig:Egs_Rch_O16_NN_3N} and \ref{fig:Egs_Rch_Ni56_NN_3N} we show
the ground-state energies and charge radii of $^{16}$O and $^{56}$Ni for the
\textit{NN}-only and \textit{NN} plus 3\textit{N} interactions as a function of the harmonic-oscillator
frequency \hw and for different model-space truncations. While we observe 
converged results at \hw $\approx 20-24 \MeV$ for $^{16}$O for \textit{NN}+3\textit{N} 
interactions, the results for $^{56}$Ni are fully converged only with respect 
to the single-particle basis \emax. Increasing the 3\textit{N} cut $E_\text{3Max}$ from 
14 to 16 still results in slight changes for energies and radii. Moreover, selecting the 
optimal frequency for extracting the charge radius of $^{56}$Ni is not as 
clear as for $^{16}$O, as the results show an unusual convergence behavior 
with \hw and the model-space truncation, which could be due to the 3\textit{N} cut 
$E_\text{3Max}$.

For all following results, we choose the frequency for extracting radii consistent 
with the ground-state energy, keeping in mind that the results for the charge 
radii of the nickel and heavier calcium isotopes are somewhat less converged.
Calculations based on \textit{NN}-only interactions are well converged for both nuclei 
and the optimal \hw is shifted to slightly larger values. Generally, we find that 
3\textit{N} interactions have a significant impact on the ground-state energies and 
charge radii. In both nuclei, $^{16}$O and $^{56}$Ni, 3\textit{N} interactions provide
repulsive contributions, leading to significantly reduced binding energies and
increased charge radii. Compared to experimental values, we find an
underbinding of about $30 \MeV$ ($200 \MeV$) for $^{16}$O ($^{56}$Ni),
whereas the charge radius of $^{16}$O turns out to be too large by about $0.2 \fm$.

\begin{figure*}[t]
\centering
\includegraphics[width=0.75\textwidth,clip=]{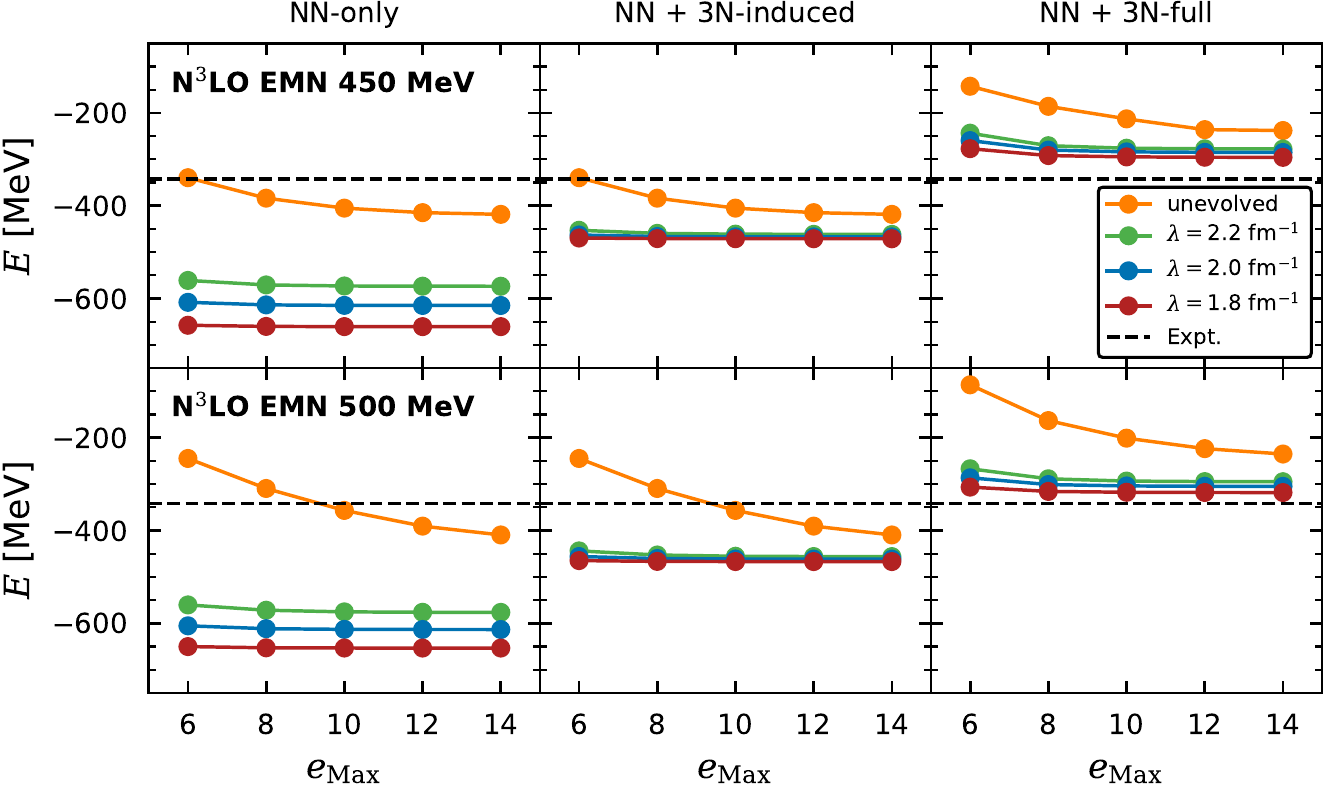}
\caption{\label{fig:Egs_SRG_Ca40}
Ground-state energy of $^{40}$Ca as a function of \emax for the \textit{NN}-only
(left), \textit{NN}+3\textit{N}-induced (middle), and \textit{NN}+3\textit{N}-full (right) interactions of
the N$^3$LO EMN 450 MeV and 500~MeV potentials, unevolved  and SRG-evolved
to resolution scales $\lambda = 2.2, 2.0,$ and $1.8 \fmi$, respectively. Results
are shown for \hw $= 20 \MeV$ and $E_\text{3Max} = 14$. The 
experimental value from Ref.~\cite{Wang17AME16} is given by the black-dashed line. }
\end{figure*}

\begin{figure*}[t]
\centering
\includegraphics[width=0.75\textwidth,clip=]{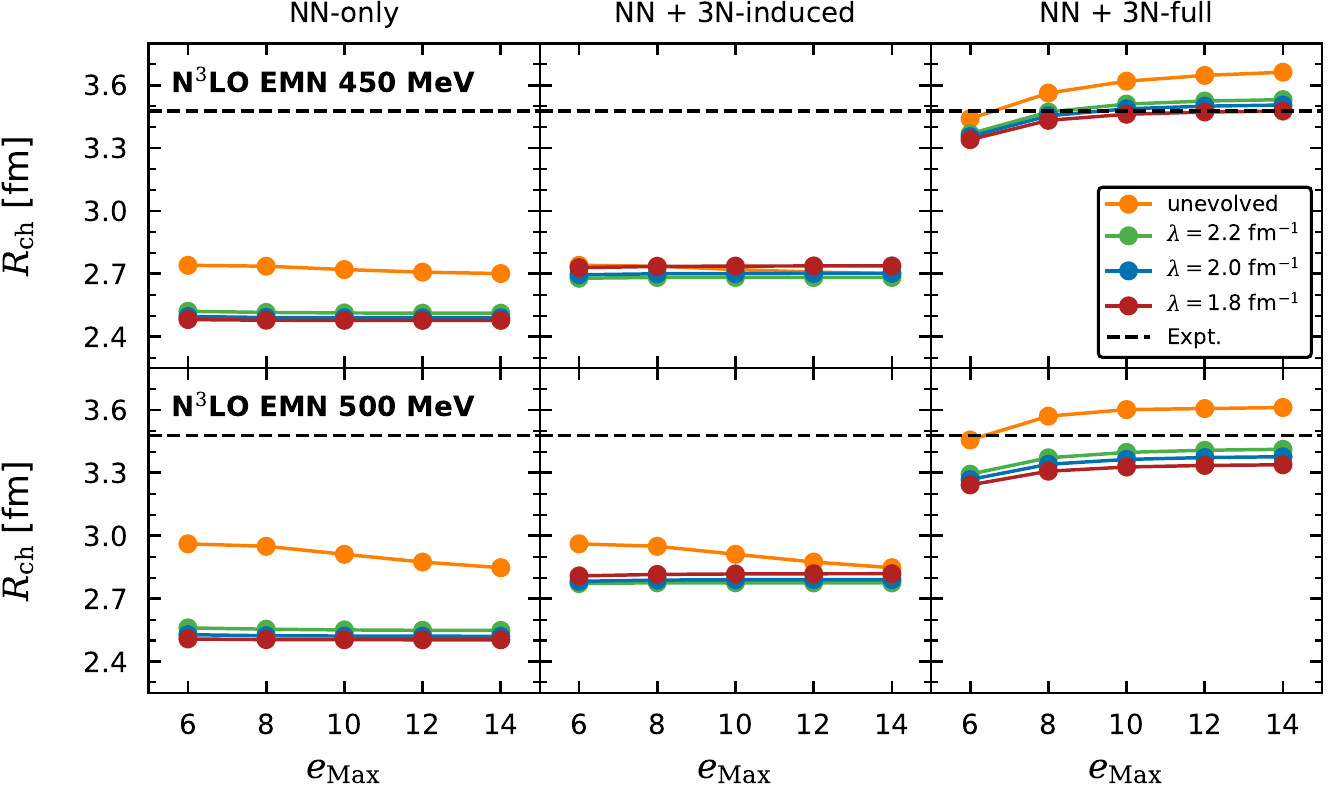}
\caption{\label{fig:Rch_SRG_Ca40}
Same as Fig.~\ref{fig:Egs_SRG_Ca40} but for the charge radius of $^{40}$Ca. The experimental
value is taken from Ref.~\cite{Ange13rch}. Note that the results for unevolved interactions are 
barely visible in some panels, as they are on top of the corresponding evolved results.}
\end{figure*} 	
 
\begin{figure*}[t!]
\centering
\includegraphics[width=0.675\textwidth,clip=]{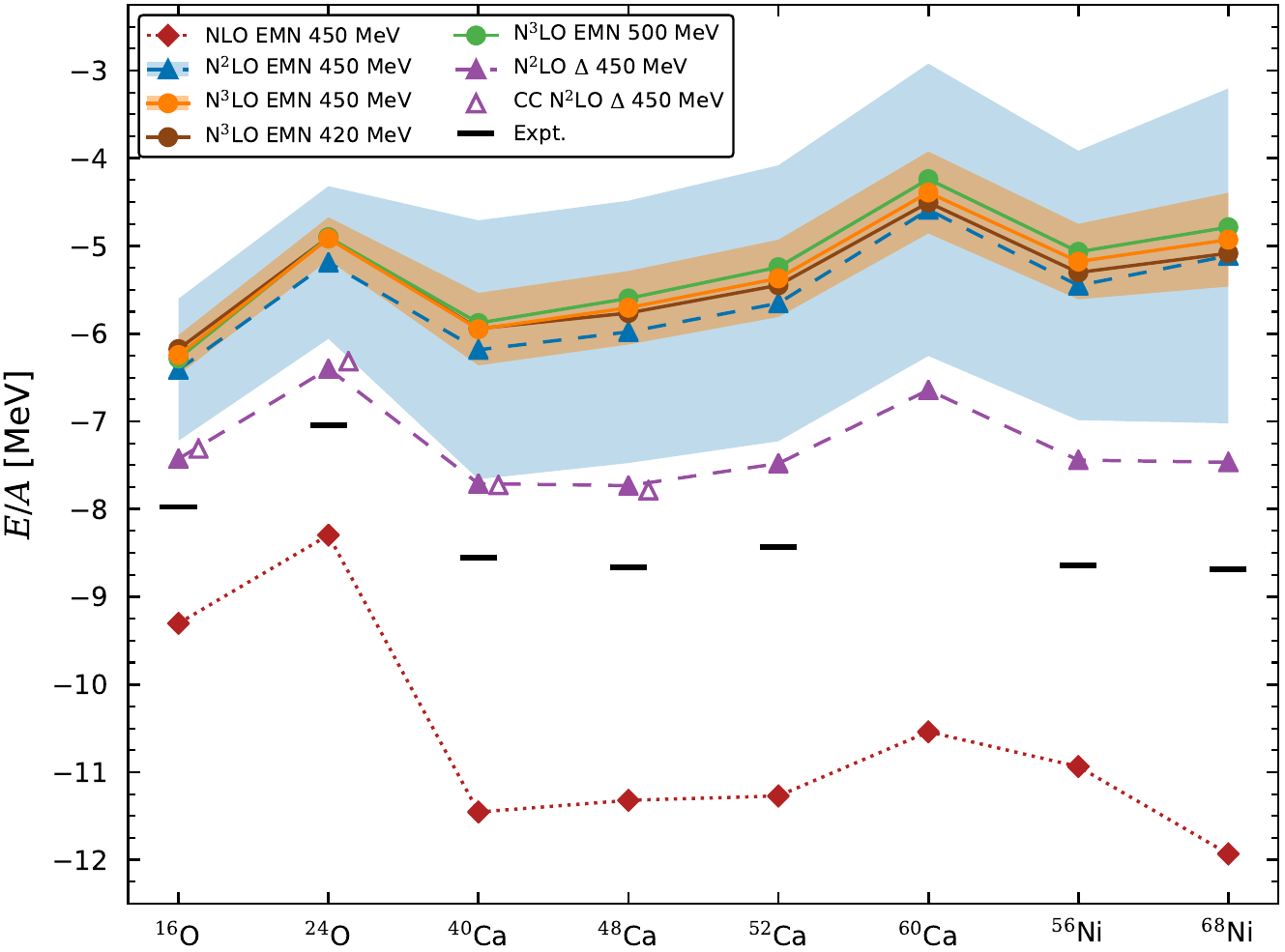}
\caption{\label{fig:Egs_closed_shell}
Ground-state energies per nucleon of selected closed-shell oxygen, calcium, and nickel
isotopes. Results are shown at N$^3$LO for the EMN potential with cutoffs $\Lambda=420, 450$,
and $500 \MeV$ depicted by the brown, orange, and green-solid lines and circles, 
respectively. The N$^2$LO results are given by the dashed lines for the EMN $450 \MeV$
potential (blue line and solid up triangles) and the $\Delta$-full interaction (purple line and solid up 
triangles), while NLO results are displayed by the red-dotted line and diamonds. The open 
triangles give the coupled cluster (CC) results for the $\Delta$-full interaction from 
Ref.~\cite{Ekst17deltasat} for comparison. The blue and orange bands give the N$^2$LO 
and N$^3$LO uncertainty estimate, respectively, for the EMN $450 \MeV$ interaction.
We note that the uncertainty due to the $E_\text{3Max}$ cut is $\lesssim 0.1 \MeV/A$ 
through $^{40}$Ca and increases up to $\sim 0.5 \MeV/A$ for $^{68}$Ni. Experimental values 
are taken from Ref.~\cite{Wang17AME16}.}
\end{figure*}

\begin{figure*}[t!]
\centering
\includegraphics[width=0.675\textwidth,clip=]{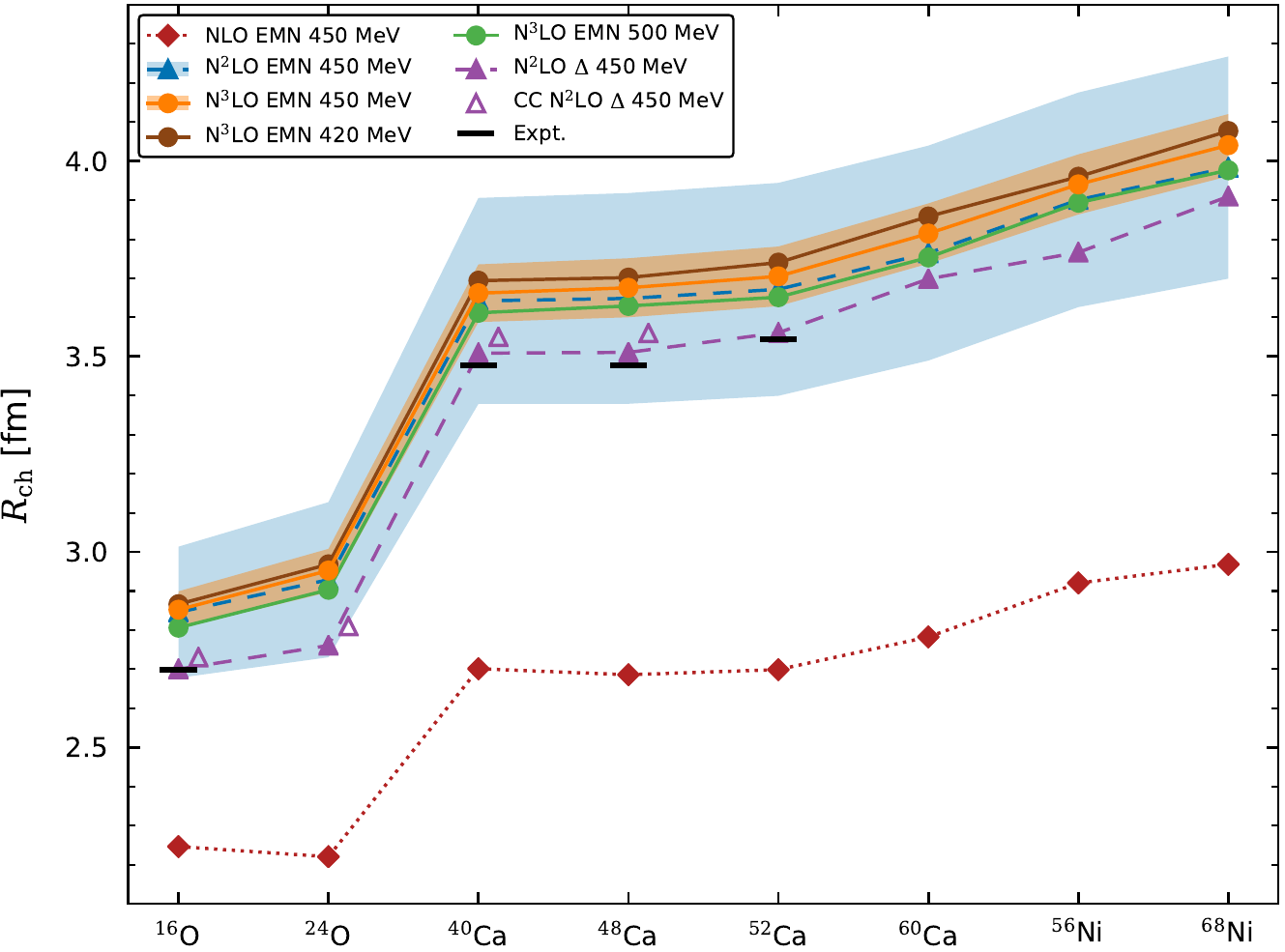}
\caption{\label{fig:Rch_closed_shell}
Same as Fig.~\ref{fig:Egs_closed_shell} but for charge radii. Experimental values are taken
from Refs.~\cite{Ange13rch,Ruiz16Calcium}. Note that the results for the heavier calcium isotopes and beyond
are somewhat less converged in \hw (see text for details).}
\end{figure*}

In the following, we study $^{40}$Ca based on consistently-evolved \textit{NN}+3\textit{N} interactions
following Ref.~\cite{Hebe12msSRG}. We distinguish three cases: ``\textit{NN}-only'' (no 3\textit{N} 
contributions at all), ``\textit{NN}+3\textit{N}-induced'' (\textit{NN} plus induced 3\textit{N} contributions from the
\textit{NN} SRG evolution in momentum space), and ``\textit{NN}+3\textit{N}-full'' (including also initial 3\textit{N} 
interactions in the SRG evolution). Our results for the ground-state energy and charge radius of $^{40}$Ca
are shown in Figs.~\ref{fig:Egs_SRG_Ca40} and~\ref{fig:Rch_SRG_Ca40} for the N$^3$LO
EMN $450 \MeV$ and $500 \MeV$ interactions. 
Note that the radius operator is not free-space 
SRG evolved.

As in Figs.~\ref{fig:Egs_Rch_O16_NN_3N} and \ref{fig:Egs_Rch_Ni56_NN_3N}, we
find significantly decreased binding energies and increased charge radii due to 
3\textit{N} interactions, also for the $500 \MeV$ interaction. The ``\textit{NN}-only'' results exhibit 
a sizable resolution scale dependence, whereas ``\textit{NN}+3\textit{N}-induced'' leads to very 
similar results in the studied range $\lambda = 1.8-2.2 \fmi$. This indicates that 
contributions from neglected induced higher-body interactions are rather insignificant.
However, although the results with unevolved interactions appear to be converged
with respect to the model space \emax for $\Lambda = 450 \MeV$, they
differ significantly from the results with evolved interactions. This indicates that 
contributions beyond the IM-SRG(2) are indeed relevant. The ``\textit{NN}+3\textit{N}-full'' 
ground-state energies are remarkably similar for both N$^3$LO
EMN $450 \MeV$ and $500 \MeV$ interactions. This is most likely due
to fitting the 3\textit{N} couplings to the same nuclear matter observables. However, 
as in Figs.~\ref{fig:Egs_Rch_O16_NN_3N} and \ref{fig:Egs_Rch_Ni56_NN_3N}, 
there are similar deficiencies with respect to experiment, with a difference
of about $2 \MeV$ per nucleon (see also Fig.~\ref{fig:Egs_closed_shell}).

We find similar trends for the charge radius in Fig.~\ref{fig:Rch_SRG_Ca40}.
In this case, the ``\textit{NN}+3\textit{N}-induced'' results are very similar for the unevolved
interaction and all resolution scales studied, indicating that contributions beyond
the IM-SRG(2) may be less relevant for this observable.
This observation could point to the missing IM-SRG(3) contributions being of
short-range character, given that the radius operator is mainly sensitive to
long-range contributions.
 We find again 
remarkably similar results for both N$^3$LO EMN $450 \MeV$ and 
$500 \MeV$ interactions. In contrast to the results for the ground-state energy
(see Fig.~\ref{fig:Egs_SRG_Ca40}), we find a better agreement with the
experimental charge radius for all ``\textit{NN}+3\textit{N}-full'' calculations (note the scale
in Fig.~\ref{fig:Egs_SRG_Ca40} compared to Fig.~\ref{fig:Rch_SRG_Ca40}).

In the following, we consider a model space of \emax/$E_{\text{3Max}} = 14/14$ and 
\hw $=20 \MeV$. In Figs.~\ref{fig:Egs_closed_shell} and~\ref{fig:Rch_closed_shell}, 
we show results for ground-state energies and charge radii of selected
closed-shell oxygen, calcium, and nickel isotopes for the (unevolved) N$^3$LO 
EMN 420, 450, and $500 \MeV$ interactions. In addition, we present results at 
NLO and N$^2$LO for the EMN $450 \MeV$ interaction. This enables us to provide
uncertainty estimates for the order-by-order convergence of the chiral expansion
(see, e.g., Ref.~\cite{Epel15improved}). For the orders $i \geqslant 3$
(i.e., $\geqslant$ N$^2$LO), the uncertainty $\Delta X^{(i)}$ for a fixed cutoff is 
estimated by
\begin{equation}
\Delta X^{(i)} = \max_{3\leqslant j \leqslant i} (Q^{i+1-j}|X^{(j)}-X^{(j-1)}|) \,,
\end{equation}
where  $X^{(j)}$ denotes the obtained result at order $j$ in chiral EFT and 
$Q = m_\pi/\Lambda_b$ is the ratio of a typical momentum scale over the 
breakdown scale, with the pion mass $m_\pi=140 \MeV$ and we take
for the breakdown scale $\Lambda_b = 500 \MeV$. In Figs.~\ref{fig:Egs_closed_shell} 
and~\ref{fig:Rch_closed_shell}, the uncertainty estimates for the EMN $450 \MeV$
interaction at N$^2$LO and N$^3$LO are depicted by the blue and orange band, 
respectively. We have assessed the convergence with respect to $E_\text{3Max}$
by increasing its value to 16 in selected cases, leading to changes of ground-state
energies up to about $0.1~\MeV/A$ until $^{40}$Ca and up to $0.5~\MeV/A$ for 
$^{68}$Ni. The changes of radii are only minor. However, as discussed above
we note that contributions beyond the IM-SRG(2) may be important and need
to also be explored explicitly in the future. For comparison, we also include
in Figs.~\ref{fig:Egs_closed_shell} and~\ref{fig:Rch_closed_shell} results based
on a recently developed $\Delta$-full interaction~\cite{Ekst17deltasat} at N$^2$LO,
using the same model space but \hw $=16 \MeV$.
This also shows the excellent comparison of our IM-SRG(2) calculations with the coupled cluster (CC) results from Ref.~\cite{Ekst17deltasat}.

The NLO interaction significantly overbinds all nuclei. Adding N$^2$LO leads
to substantial repulsive contributions, resulting in an underbinding compared
to experiment. The impact of N$^3$LO on the ground-state energies is rather 
small. Overall the results exhibit a systematic order-by-order convergence
with overlapping N$^2$LO and N$^3$LO bands. Moreover, we observe only 
a weak cutoff dependence at N$^3$LO, which slightly increases for larger 
mass numbers, and the results for all cutoffs are within the (orange) uncertainty
band. The underbinding compared to experiment is expected from the comparison
of the N$^3$LO EMN $450 \MeV$ interaction with the empirical saturation
region (see Fig.~4 in Ref.~\cite{Dris17MCshort}). The ground-state energies 
resulting from the $\Delta$-full interaction are in better agreement with 
experiment, but still underbind the investigated closed-shell nuclei (see 
also Ref.~\cite{Ekst17deltasat}).

\begin{figure}[t]
\centering
\includegraphics[width=\columnwidth,clip=]{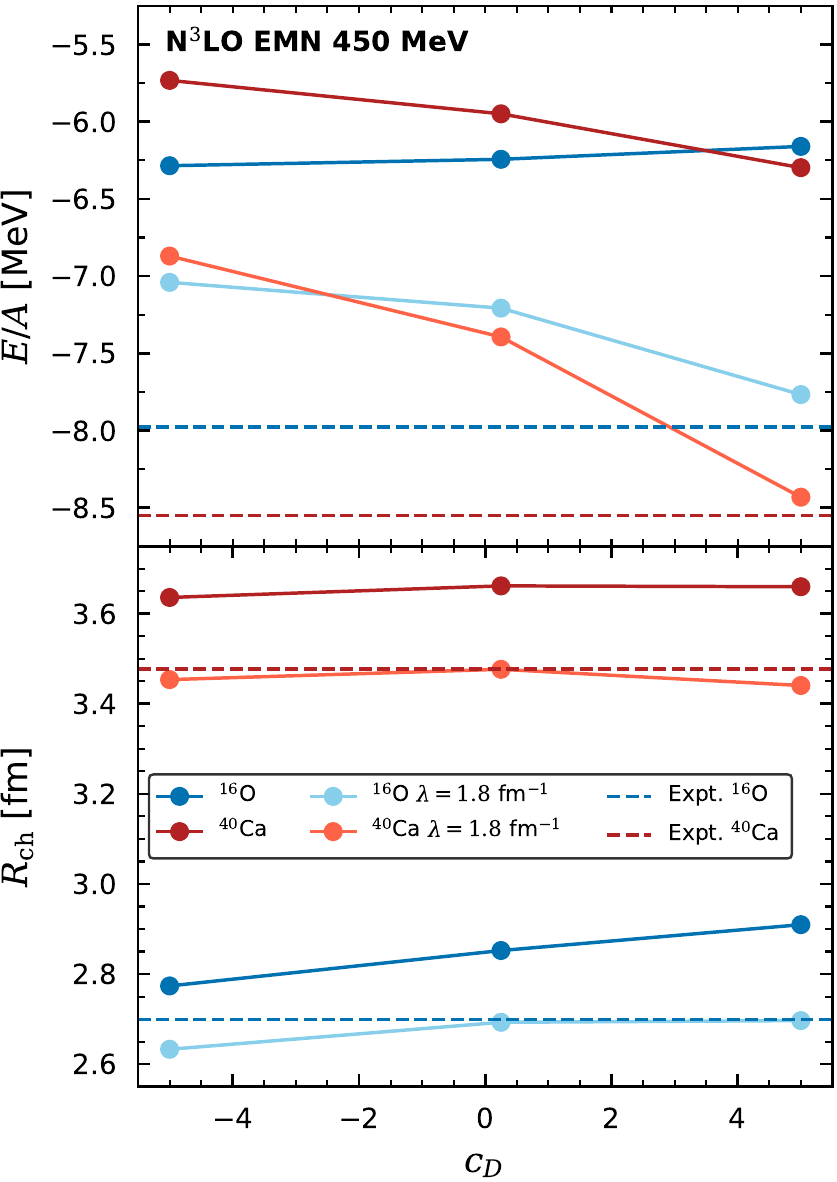}
\caption{\label{fig:Egs_Rch_cD_cE_var}
Ground-state energies (top) and charge radii (bottom panel) for $^{16}$O and $^{40}$Ca 
as a function of $c_D$ (with corresponding $c_E$ value from the triton binding energy, see 
Table~\ref{tab:cD_cE}). Results are shown for unevolved and SRG-evolved potentials 
with $\lambda=1.8 \fmi$. The experimental values from Refs.~\cite{Wang17AME16,Ange13rch}
are given by the dashed lines. }
\end{figure}

\begin{figure}[t]
\centering
\includegraphics[width=\columnwidth,clip=]{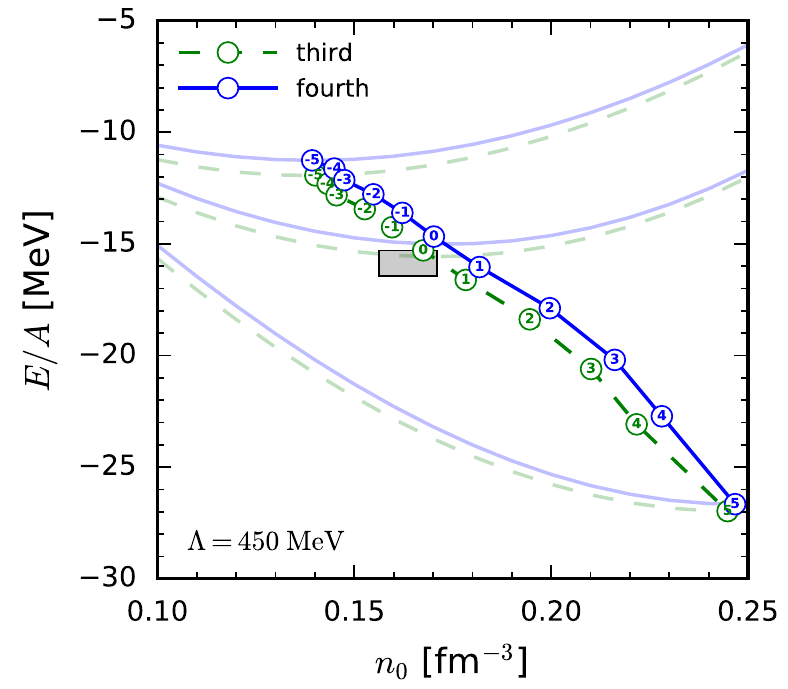}
\caption{\label{fig:cD_extended_EMN_450}
Same as Fig.~\ref{fig:cD_cE_EMN400_420} but for the N$^3$LO EMN $450 \MeV$
interaction. Note the increased range in density and in $c_D$ couplings (annotated).
We also show the energy per particle for three interactions with $c_D = -5, 0.25,$ 
and $5$ at third and fourth order in MBPT. For $c_D = 5$, the saturation
point is more exploratory, as this is not as constrained in density
from our calculations up to densities of $0.25 \fmiq$.}
\end{figure}

\begin{figure}[t]
\centering
\includegraphics[width=\columnwidth,clip=]{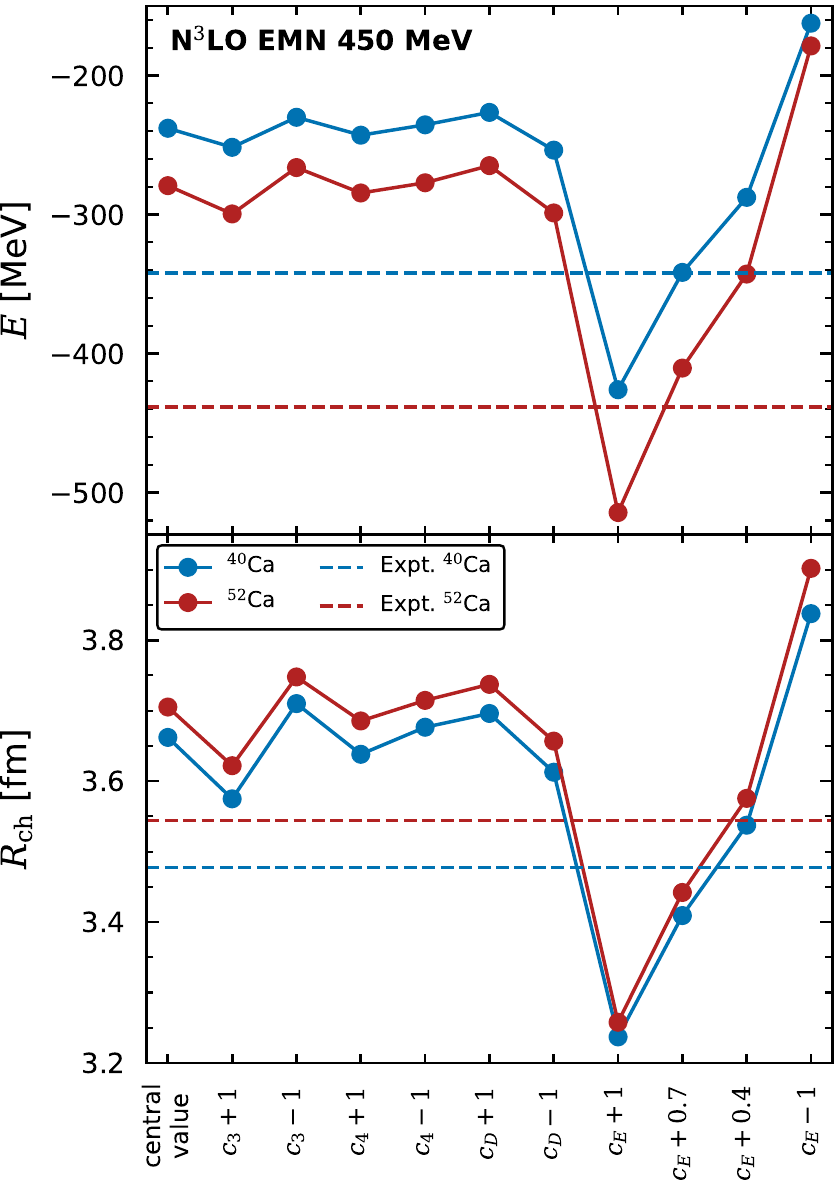}
\caption{\label{fig:Egs_Rch_ci_var_Ca}
Ground-state energies (top) and charge radii (bottom panel) for $^{40}$Ca and
$^{52}$Ca for variations of the long-range 3\textit{N} couplings $c_3$, $c_4$ by $\pm 1 \, \text{GeV}^{-1}$
and the shorter-range 3\textit{N} couplings $c_D$, $c_E$ by $\pm 1$. Varying $c_1$ by $\pm 1 \, \text{GeV}^{-1}$
leads to similar results as for $c_3$. In addition, we show results for $c_E+0.7$ 
and $c_E+0.4$. The first point (central value) is for the original 
N$^3$LO EMN $450 \MeV$ interaction. The experimental values from 
Refs.~\cite{Wang17AME16,Ange13rch,Ruiz16Calcium} are given by the dashed lines.}
\end{figure}

\begin{figure}[t]
\centering
\includegraphics[width=\columnwidth,clip=]{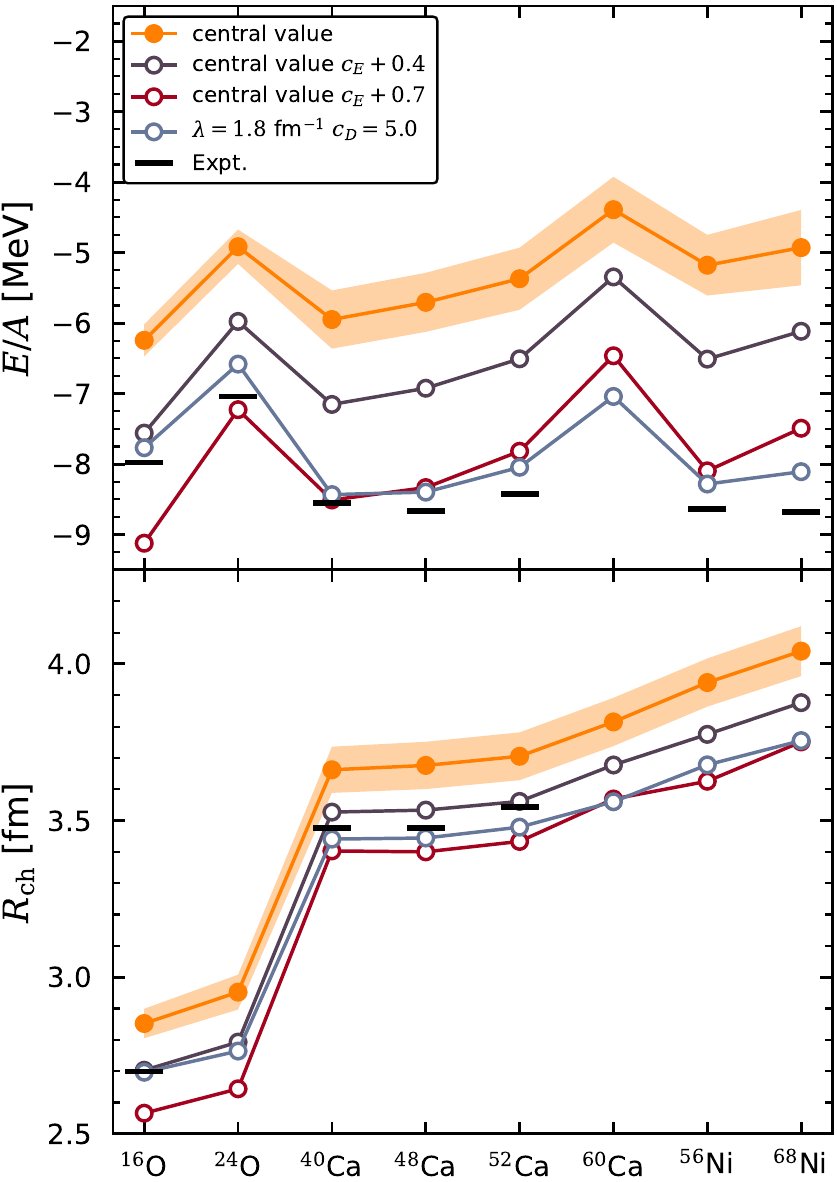}
\caption{\label{fig:Egs_Rch_closed_shell_cE}
Ground-state energies per nucleon and charge radii of selected closed-shell oxygen,
calcium, and nickel isotopes for the original N$^3$LO EMN $450 \MeV$ interaction
(central value), the modified $c_E+0.4$ as well as $c_E+0.7$ interaction,
and the SRG-evolved interaction with $\lambda=1.8 \fmi$ (with $c_D=5.0$).
The orange band represents the N$^3$LO uncertainty estimate from
Figs.~\ref{fig:Egs_closed_shell} and~\ref{fig:Rch_closed_shell}. Experimental
values are taken from Refs.~\cite{Wang17AME16,Ange13rch,Ruiz16Calcium}.}
\end{figure}

The general trends for the charge radii are similar to the observations for
the ground-state energies, with systematic order-by-order convergence,
overlapping uncertainty bands, and small N$^3$LO cutoff variation. Overall
we find too large radii, but the N$^2$LO uncertainty band encloses the 
experimental values, while the $\Delta$-full interaction again exhibits better
agreement with experiment. The correlation with the empirical saturation 
region needs further studies in this case, as one would have expected
smaller radii based on nuclear matter saturation for the N$^3$LO EMN 
$450 \MeV$ interaction (see again Fig.~4 in Ref.~\cite{Dris17MCshort}).

Our results for the ground-state energies and charge radii indicate that 
a realistic description of only the saturation point of nuclear matter may
not be sufficient for a realistic description of medium-mass nuclei. To 
shed more light on this, we study the sensitivity of our results on variations 
of the 3\textit{N} couplings $c_D$ and $c_E$, constrained only by fits to the 
triton binding energy (see Table~\ref{tab:cD_cE}), without the constraint
to the empirical saturation region. The results for the ground-state 
energies and charge radii of $^{16}$O and $^{40}$Ca as a
function of $c_D$ are shown in Fig.~\ref{fig:Egs_Rch_cD_cE_var}.
Although the saturation point of nuclear matter is very sensitive to the
values of $c_D$, as shown by the large variation in Fig.~\ref{fig:cD_extended_EMN_450},
the variation of the ground-state energy of $^{16}$O and $^{40}$Ca
over the range of $c_D = -5 \ldots 5$ is much smaller, which points
to lower nuclear matter densities being more relevant, and the variation
of the charge radii is very small. We have checked that the weaker
impact of $c_D$ also persists for the other studied cutoffs at N$^3$LO.
For illustration, Fig.~\ref{fig:cD_extended_EMN_450} also shows 
the nuclear matter energy per particle at third and fourth order in MBPT for three 
$c_D$ values ($-5, 0.25, 5$)\footnote{Note that the minima of the nuclear
matter energy curves can be slightly different from the approximate
$c_D$ fitting procedure, such that the saturation points may not be exactly 
the same for the curves and circles in Fig.~\ref{fig:cD_extended_EMN_450}.}.
Tracing these to lower densities around $0.1 \fmiq$ or below gives an energy 
range for the $c_D$ variation, which is 
much smaller than at saturation density. Even though the change in energy is still larger than for finite nuclei, the lower densities resemble more closely the results for 
$^{16}$O and $^{40}$Ca.
 The sensitivity of the ground-state 
energies in Fig.~\ref{fig:Egs_Rch_cD_cE_var} to $c_D$  increases for 
the SRG-evolved interactions, but is still much smaller than for nuclear 
matter at saturation density. More work is thus needed to establish in
which way nuclear matter properties are most constraining for the
development of novel nuclear forces that lead to accurate results
for medium-mass and heavy nuclei.

To explore more comprehensively how sensitive our results are to the
3\textit{N} couplings, we vary their values independently, starting from the
N$^3$LO EMN $450 \MeV$ interaction. In Fig.~\ref{fig:Egs_Rch_ci_var_Ca},
we present the ground-state energies and charge radii of the calcium
isotopes $^{40}$Ca and $^{52}$Ca for variations of the
long-range 3\textit{N} couplings $c_3$, $c_4$ by $\pm 1 \, \text{GeV}^{-1}$
and the shorter-range 3\textit{N} couplings $c_D$, $c_E$ by $\pm 1$. 
Varying $c_1$ by $\pm 1 \, \text{GeV}^{-1}$ leads to similar results as for $c_3$. 
This exploratory study thus ignores correlations among these low-energy couplings (see, e.g., Ref.~\cite{Hofe15piNchiral}).
All of our variations have a relatively small impact on energies and radii,
apart from variations in $c_E$, which lead to significant changes of
$\Delta E \approx 260 \MeV$ and $\Delta R_\text{ch} \approx 0.6 \fm$
for $^{40}$Ca and similarly for $^{52}$Ca. We therefore consider two 
additional variations, $c_E+0.4$ and $c_E+0.7$. The latter reproduces 
well the ground-state energy of $^{40}$Ca and leads to an improvement
for $^{52}$Ca. We also found that setting all 3\textit{N} couplings but the
$c_3$ contribution to zero leads to similar results as the original
N$^3$LO EMN $450 \MeV$ interaction.

\begin{table}[t]
\caption{$^3$H ground-state energy and charge radius for the 
N$^3$LO EMN $450 \MeV$ interaction with modified 3\textit{N} couplings.}
\begin{ruledtabular}
\begin{tabular}{cccccc} 
& \hspace{0.05cm} Expt. \hspace{0.05cm}
& \hspace{0.05cm} $c_E + 0.4 $ \hspace{0.05cm} 
& \hspace{0.05cm} $c_E + 0.7$ \hspace{0.05cm} 
& \hspace{0.05cm} $c_3 + 1$ \hspace{0.05cm} 
& \hspace{0.05cm} $c_4 +1$ \hspace{0.05cm} \\ \hline 
$E$ [MeV] & $-8.48$ & $-9.81$ & $-11.18$ & $-8.40$ & $-8.74$ \\
$R_{\text{ch}}$ [fm] & $1.575$ & $1.469$ & $1.381$ & $1.576$ & $1.550$ \\
\end{tabular}
\end{ruledtabular}
\label{tab:Egs_Rch_triton}
\end{table}

In Fig.~\ref{fig:Egs_Rch_closed_shell_cE} we show results for selected
closed-shell nuclei exploring some of these variations compared to the
original N$^3$LO EMN $450 \MeV$ interaction (labeled central value).
The interaction $c_E + 0.7$ leads to increased binding and a better 
description of ground-state energies, whereas $c_E + 0.4$ leads to 
improved agreement with experimental charge radii. However, despite
these promising results for medium-mass nuclei, the modifications of 
the 3\textit{N} couplings lead to a heavily overbound $^3$H with a far too small 
charge radius, see Table~\ref{tab:Egs_Rch_triton}. This shows the
difficulty in achieving a realistic simultaneous description of few-body systems, 
medium-mass nuclei, and nuclear matter just by varying the 3\textit{N} couplings
based on this set of \textit{NN} interactions, and also emphasizes the need
for reliable theoretical uncertainties. In addition, we show in 
Fig.~\ref{fig:Egs_Rch_closed_shell_cE} results for the SRG-evolved
interaction with $\lambda = 1.8 \fmi$ (with $c_D = 5.0$), for which we
find ground-state energies and charge radii of $^{16}$O and $^{40}$Ca
in good agreement with experiment (see Fig.~\ref{fig:Egs_Rch_cD_cE_var}).
Figure~\ref{fig:Egs_Rch_closed_shell_cE} shows that this interaction
is also able to reproduce energies and radii of other closed-shell nuclei
in good agreement with experiment (considering the N$^3$LO uncertainties).
By construction, it still reproduces the experimental triton binding energy.

\section{Summary and outlook}
\label{sec:summary}

Our work presents the first \textit{ab initio} calculations of medium-mass nuclei
with \textit{NN}+3\textit{N} interactions to N$^3$LO. We have studied in detail ground-state 
energies and charge radii of closed-shell nuclei up to nickel for unevolved as 
well as momentum-space SRG-evolved \textit{NN}+3\textit{N} interactions with reasonable
saturation properties. In addition, we have explored a consistent momentum-space 
SRG evolution of these \textit{NN}+3\textit{N} interactions.
In general, the ground-state energies predicted by the employed interactions
significantly underbind all oxygen, calcium, and nickel isotopes studied here 
and lead to too large charge radii. Remarkably, the results exhibit only a weak
dependence on the cutoff scale. The uncertainty estimates at N$^2$LO and 
N$^3$LO for the order-by-order convergence of the chiral expansion show a 
systematic behavior with the N$^2$LO band enclosing the N$^3$LO band 
for all studied nuclei. For comparison, we also employed the $\Delta$-full
interaction of Ref.~\cite{Ekst17deltasat} at N$^2$LO, which gives improved
agreement with experimental binding energies and charge radii.

While underbinding was expected from the saturation point of the 
corresponding interactions, the behavior of the charge radii and
their correlation with the saturation point did not systematically follow 
nuclear matter. For a more detailed study of this correlation, we
varied the 3\textit{N} couplings under the constraint that the $^3$H binding energy 
agrees with experiment, and studied the resulting sensitivity of 
observables in both systems. While the ground-state energies of 
$^{16}$O and $^{40}$Ca changed by $< 1 \MeV$ for unevolved interactions, 
the change in saturation energy was $15 \MeV$ over the studied 
$c_D$ range. This indicated that nuclear matter physics at lower densities
is also relevant. When the $^3$H constraint was relaxed, we found
the largest impact on energies and radii from changes of the short-range 
3\textit{N} coupling. However, while these variations allowed us to construct
\textit{NN}+3\textit{N} interactions that lead to an improved description of medium-mass 
nuclei, the resulting nuclear forces substantially overbind the triton
and underestimate its charge radius. These findings show that the 
connection between light nuclei, medium-mass nuclei, and nuclear 
matter is quite intricate and requires further investigations. An improved
understanding of this is especially relevant for the derivation
and construction of accurate next-generation \textit{NN} and many-body 
interactions in chiral EFT.

\begin{acknowledgments}

We thank K.~Vobig for useful discussions. We also thank R.~Machleidt and
A.~Ekstr{\"o}m for providing us with the EMN and $\Delta$-full potentials,
respectively, as well as S.~R.~Stroberg for discussions on the IM-SRG
code~\cite{Stro17imsrggit}. This work was supported by the Deutsche
Forschungsgemeinschaft (DFG, German Research Foundation) -- Projektnummer
279384907 -- SFB 1245, the BMBF under Contract No.~05P18RDFN1,
the US Department of Energy, the Office of Science, the Office  of Nuclear Physics, 
and SciDAC under awards DE\_SC00046548 and DE\_AC02\_05CH11231, and by the Cluster of Excellence ``Precision Physics, Fundamental Interactions, and Structure of Matter'' (PRISMA$^+$ EXC 2118/1) funded by DFG within the German Excellence Strategy (Project ID 39083149).
C.D. acknowledges support by the Alexander von Humboldt
Foundation's Feodor-Lynen Fellowship program. Computational resources 
have been provided by the Lichtenberg high performance computer 
of the TU Darmstadt.

\end{acknowledgments}

\bibliographystyle{apsrev4-1}
\bibliography{strongint}

\end{document}